\documentclass[12pt]{article}
 
\usepackage{epsfig}
\begin{document}

\newcommand{\lfrac}[2]{\mbox{${#1}\over{#2}$}}

\baselineskip=14pt plus 0.2pt minus 0.2pt
\lineskip=14pt plus 0.2pt minus 0.2pt


\begin{center}
\large{\bf Coherent States for Unusual Potentials
} 
 
\vspace{0.25in}

\normalsize
\bigskip

Michael Martin Nieto\footnote{\noindent  Email:  
mmn@lanl.gov}\\
{\it Theoretical Division (MS-B285), Los Alamos National Laboratory\\
University of California\\
Los Alamos, New Mexico 87545, U.S.A. \\}

\normalsize

\vskip 20pt
\today

\vspace{0.3in}

{ABSTRACT}

\end{center}

\begin{quotation}
\baselineskip=.33in

The  program to construct minimum-uncertainty coherent states for 
general potentials works transparently with solvable analytic potentials. 
However, when an analytic potential is not completely solvable, 
like for a  double-well or the
linear (gravitational) potential,  there can be a conundrum.  
Motivated by supersymmetry concepts in higher dimensions, 
we show how these conundrums can be overcome.

\vspace{0.25in}

\end{quotation}

\newpage

\baselineskip=.33in

\section{Background}

The  Minimum-Uncertainty Method to obtain coherent states for general
potentials harks back to  
Schr\"odinger's discovery of the coherent states 
(of the harmonic oscillator) \cite{12}.  
It has been applied to general Hamiltonian 
potential systems, to obtain both generalized coherent states 
and generalized squeezed states \cite{n1,n2}.  

One starts with the classical Hamiltonian problem in terms of $x_c$ and
$p_c$, the classical position and momentum.  Then one transforms it into
the ``natural classical variables," $X_c$ and $P_c$, which vary as the
$\sin$ and the $\cos$ of the classical $\omega t$.  The classical
Hamiltonian is therefore of the form $P_c^2 + X_c^2 $.  
These natural classical variables are next changed into
``natural'' quantum operators.  These quantum operators have a
commutation relation and an associated uncertainty relation:  
\begin{equation} 
[X,P] = iG, \hspace{0.5in} 
(\Delta X)^2(\Delta P)^2 \geq {\frac{1}{4}}\langle G\rangle ^2, 
\label{uncert}
\end{equation}
where in general $G$ is an operator.  
The states that minimize this 
uncertainty relation are given by the solutions to the equation 
\begin{equation}
Y\psi_{ss} \equiv
\left(X + \frac{i\langle G\rangle }{2(\Delta P)^2} P\right)\psi_{ss}
=\left(\langle X\rangle +\frac{i\langle G\rangle }{2(\Delta P)^2}\langle
P\rangle 
\right)\psi_{ss}.
\end{equation}
Note that of the  four parameters $\langle X\rangle , \langle P\rangle , 
\langle P^2\rangle $, and $\langle G\rangle $, only three are
independent because they satisfy the equality in the uncertainty relation.
Therefore,
\begin{equation}
\left(X + iB P\right)\psi_{ss} = C \psi_{ss}  ,~~~
B = \frac{\Delta X}{\Delta P},  ~~~
 C = \langle X\rangle + i B \langle P\rangle .
\end{equation}
Here $B$ is real and $C$ is complex.  These states, $\psi_{ss}(B,C)$, are the 
minimum-uncertainty states for general potentials
\cite{n1,n2}.  Using later parlance, they  are  the squeezed states for
general potentials.  $B$ can be adjusted to $B_0$ so that the ground
eigenstate of the potential is a member of the set.  Then these restricted
states, $\psi_{ss}(B=B_0,C)=\psi_{cs}(B_0,C)$, 
are the minimum-uncertainty coherent states for general  potentials.

It can be understood that $\psi_{ss}(B,C)$ and
$\psi_{ss}(B_0,C)$ are  the squeezed and coherent 
states by recalling the situation for the harmonic oscillator.  
The harmonic oscillator coherent states are the displaced ground state.  
The harmonic oscillator squeezed states are
Gaussians that have  widths different than that of the ground state 
Gaussian, which are then displaced. 

For the harmonic oscillator these coherent states are equivalent to those 
obtained from the ladder operator method:
\begin{equation}
a|\alpha\rangle = \alpha|\alpha\rangle.   
\end{equation}
In general the $X$ and $P$ operators can be given in terms 
of the raising and lowering operators (or their $n$-dependent
generalizations):    
\begin{equation}
X= \frac{1}{\sqrt{2}}[a + a^\dagger], ~~~~~~~
P= \frac{1}{i\sqrt{2}}[a - a^\dagger].
\end{equation}
Here $a$ and $a^\dagger$ are the lowering and raising operators of the
system.


\section{The double-well and linear potential conundrums}

Although this procedure works well for exactly solvable systems, one nagging
question has always been if one could, in principle, handle double-well
potentials.  This question was raised by a number of 
people, in particular by Rohrlich  \cite{barut}. The problem was that no
completely solvable double-well potential existed.  
Therefore, earlier techniques 
could not give a demonstration that a coherent-state procedure could 
analytically work for a double well.  In the following Section \ref{susyV} 
we discuss supersymmetry techniques that have now been developed and then 
apply them to a double-well system in Section  \ref{dwcs}. 

Another specific problem has to do with the linear potential, $V=mg|x|$.  
As was pointed out by Kienle and Straub \cite{ks}, the standard 
method of  solution for the natural classical variables \cite{n1,n2} 
breaks down here.   Usually,  solving for the natural variables 
amounts to solving the differential equation
\begin{equation}
\frac{d}{dx}X_c(x) = \omega_c(E_c)\left(\frac{m}{2}\right)^{1/2}
\left[\frac{X_{c(MAX)}^2-X_c^2(x)}{E_c-V(x)}\right]^{1/2}. \label{mucssolve}
\end{equation}
For normal systems, like $V \propto \{x^2,~-1/\cosh^2x\}$, respectively,
Eq. (\ref{mucssolve}) is easily solved; $X_c \propto \{x,~\cosh x\}$, 
respectively.  Here
things get singular.  In Section \ref{lincs} we apply the same 
supersymmetric techniques of Section \ref{susyV} to resolve this problem.  


\section{Supersymmetry for Volcano Potentials}
\label{susyV}

We first  remind the reader of another type of uncommon potential, 
volcano-shaped potentials.     
They turn out to be of current interest in theories of higher
dimensions \cite{arkani}-\cite{hill}.  
In these theories  one can be trying to discover if 
volcano-shaped  potentials have  zero-energy bound states 
satisfying supersymmetry \cite{mmns}, in what amounts to 
a 1-dimensional Schr\"odinger equation \cite{mmnvolc}.   

\begin{figure}[h]
 \begin{center}
\noindent    
\psfig{figure=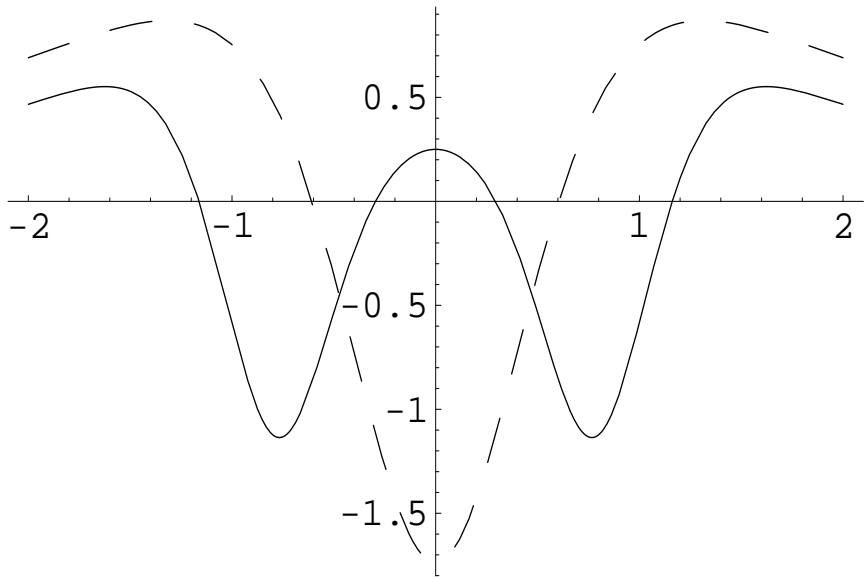,width=4.5in,height=3in}
\caption{  
The dashed and solid curves show the potentials of Eqs. (\ref{mmnsusyV})
and (\ref{Vplug}), respectively, both of which are supersymmetric and have 
zero-energy ground states.  
\label{vplugv}}
\end{center}
\end{figure}

An example is the volcano potential  
(shown in Figure \ref{vplugv})
\begin{equation}
V(z) = \frac{-\left(\sqrt{5} - \frac{1}{2}\right) +\frac{19}{4}z^2}
           {[1+z^2]^2}.  
\label{mmnsusyV}
\end{equation} 


\noindent
(For the rest of this paper we use the unitless quantities defined by 
$\{\hbar,~m\}\rightarrow 1$, with factors such as $2$ absorbed into  
$x\rightarrow z$. Thus, the first term in the  Schr\"odinger equation will
always have the form $-\partial^2$).

Equation (\ref{mmnsusyV})  is a (Schr\"odinger-like factorization) 
supersymmetric potential \cite{mmns} of the form
\begin{eqnarray}
V(z) &=& \left[ W'(z)\right]^2 - W''(z).  \label {vasw} \\
\psi_0(z)&=&N \exp[-W(z)],  \label{psiasw} \\
W(z)&=&\left[\frac{\sqrt{5}}{2}-\frac{1}{4}\right] 
       \ln\left(1 + z^2\right).  \label{susyvl}
\end{eqnarray} 
The Hamiltonian can be written as 
\begin{eqnarray}
H &=& -\partial^2 +V(z) = A^\dagger A, \\
A &=& \partial + W'(z). 
\end{eqnarray}

Any potential that is supersymmetric 
has a  ground state with zero-energy. 
Sometimes a constant must be added to a potential to make it  
supersymmetric.  For example, the hydrogen atom and simple harmonic
oscillator potentials can be made supersymmetric by subtracting the
original ground-state energies from the potentials \cite{HOHA}.  

But playing with the form of the above $W(z)$, one can quickly
convince oneself that various shaped potentials can be obtained.  
For example, a volcano potential with a plug in the center is given by 
(also shown in Figure \ref{vplugv})
\begin{eqnarray}
W(z)&=&\lfrac{1}{4}\ln\left[1-\frac{1}{2}x^2+x^4\right], \\
V(z)&=&\frac{{1-\lfrac{45}{4}x^2-3x^4+8x^6}}
{4\left(1-x^2/2+x^4\right)^2}.   
\label{Vplug}
\end{eqnarray}

Note that although we have analytic potentials and 
analytic ground states, here we do not have the excited spectra 
and their wave functions.  However, because of the properties 
of supersymmetry, it will turn out that such a 
situation will be sufficient to resolve our conundrums.  


\section{Demonstrating coherent states for a double-well potential}
\label{dwcs}

\begin{figure}[h]
 \begin{center}
\noindent    
\psfig{figure=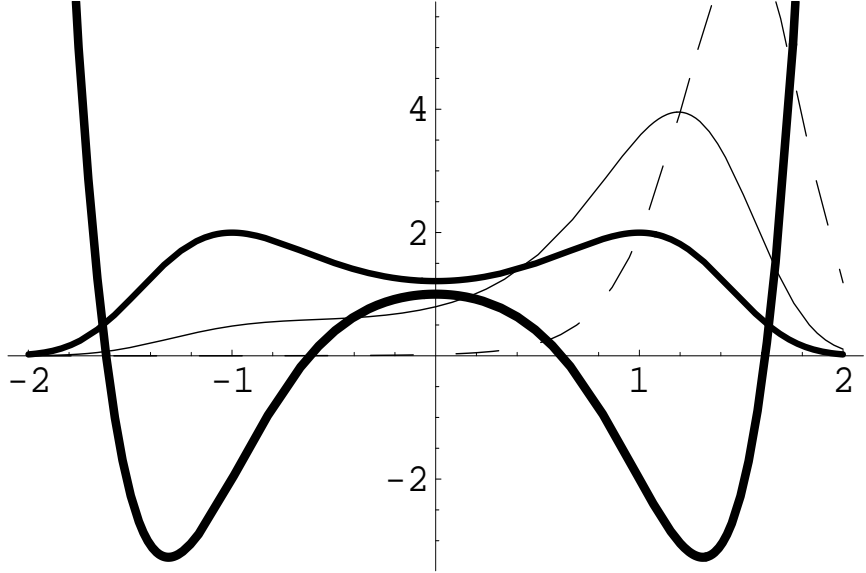,width=4.5in,height=3in}
\caption{  
The thick curve plots the potential of Eq. (\ref{dwV}).  To maintain the 
same scale, 5 times the  mod-squares of various coherent-state wave
functions are shown.  The  
ground-state wave function is given by the medium-thick 
double-humped curve. The coherent state with $\alpha=1/2$ is the thin
curve, and the coherent state with  $\alpha=2$  is the dashed curve. 
\label{dwVandpsi}}
\end{center}
\end{figure}

Stimulated by the results of the last section, consider the function  
\begin{equation}
W(z) = -\lfrac{1}{2}z^2+\lfrac{1}{4}z^4.
\end{equation}
This yields the supersymmetric potential 
\begin{equation}
V(z) = 1-2z^2-2z^4+z^6  \label{dwV}
\end{equation} 
with normalized zero-energy ground state wave function
\begin{equation}
\psi_0(z) = N_0\exp[-W(z)] =
[2.0410\dots] \exp\left[\lfrac{1}{2}z^2-\lfrac{1}{4}z^4\right].
\label{dwpsi}
\end{equation}
These quantities are shown in Figure \ref{dwVandpsi}.  In particular, 
as it should, the ground state
wave packet has no zero but double humps centered at the potential's
minima.  

Now consider the supersymmetry (factorized) annihilation operator 
for this system
\begin{equation}
A(z) =  \partial_z + W'(z).  
 \end{equation}
Using this in the ladder-operator definition of coherent states,
\begin{equation}
A(z) \psi_\alpha(z)=\alpha~ \psi_\alpha(z)
\end{equation}
yields 
\begin{equation}
\psi_\alpha(z)=N_\alpha\exp\left[\alpha z-W(z)\right]. \label{psialpha}
\end{equation}

In Figure \ref{dwVandpsi} we also show the coherent-state wave packets for
$\alpha=\{1/2,2\}$.  As $\alpha$ becomes larger the wave packet moves
further to the right and assumes a more peaked form, as coherent states
should.  (For negative $\alpha$ the parity-reversed situation occurs.) 
Eq.  (\ref{psialpha}) shows that these states partially resemble 
displacement-operator states. Also, if one defines 
one's $X$ and $P$ in terms of the sums and differences of the
$A$ and $A^\dagger$ then they also have a minimum-uncertainty
characteristic.  So, these coherent states obtained from
supersymmetry/factorization are well behaved.  This is because $A$ is the
ground-state annihilation operator.


\section{Coherent states for the linear (gravitational) potential}
\label{lincs}

\begin{figure}[h]
 \begin{center}
\noindent    
\psfig{figure=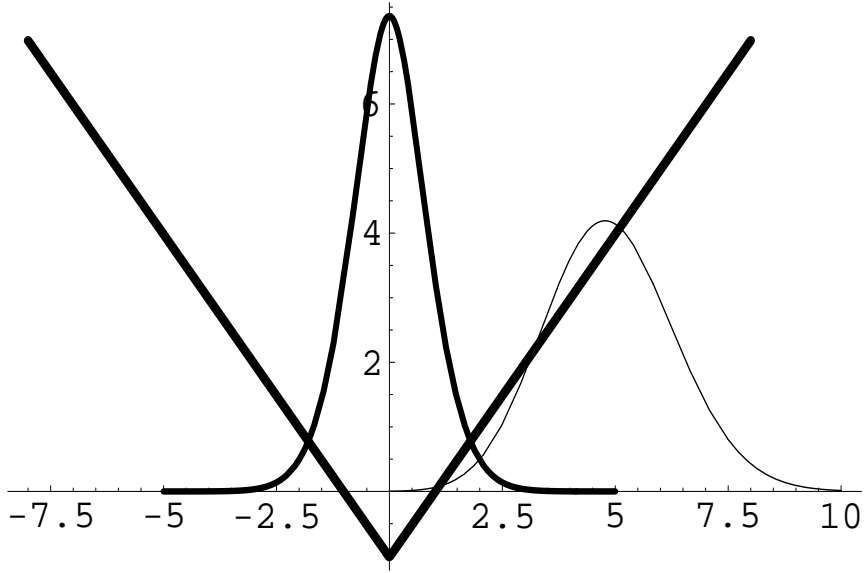,width=4.5in,height=3in}
\caption{  
The thick curve show the supersymmetric linear potential.  To maintain the 
same scale, 15 times the  mod-squares of various coherent-state wave
functions are shown.  The  
ground-state wave packet is given by the medium-thick 
curve. The coherent state with $\alpha=2$ is the thin curve. 
\label{linVandpsi}}
\end{center}
\end{figure}

Using units where $(2m^2g/\hbar^2)^{1/3}x \rightarrow z$, the Schr\"odinger
equation is 
\begin{equation}
\left[-\frac{d^2}{dz^2} + |z|\right]\psi_n(z) = \lambda_n\psi_n(z).
\end{equation}  
This is Airy's equation and, with foresight, we subtract off the ground
state eigenenergy to make the system supersymmetric: 
\begin{equation}
\left[-\frac{d^2}{dz^2} + (|z|-\lambda_0)\right]\psi_n(z) 
= (\lambda_n-\lambda_0)\psi_n(z)\equiv \Lambda_n\psi_n(z).
\end{equation}
$\lambda_0= 1.018...$. It and the other $\lambda_n$ are well known
numerically \cite{airynumer}.  The ground state-solution is the Airy 
function \cite{linsolve}
\begin{equation}
\psi_0 = N_0~Ai(|z|-\lambda_0). 
\end{equation}
In Figure \ref{linVandpsi} we show the supersymmetric potential and the
zero-energy ground state wave packet.

But now we can give the supersymmetric function as simply being 
\begin{equation}
W(z) = -\ln\left[Ai(|z|-\lambda_0)\right],
\end{equation}
and the formalism follows through.  This means we can write the coherent
states as 
\begin{eqnarray}
A(z)\psi_\alpha(z) &=& \left[\partial_z + W'(z)\right] \psi_\alpha(z)
= \alpha~\psi_\alpha(z), \\
\psi_\alpha(z)&=&N_\alpha\exp[\alpha z]Ai(|z|-\lambda_0). 
\label{psialphalin}
\end{eqnarray}
In Figure \ref{linVandpsi} we also shown the coherent state for
$\alpha=2$. 

Once again the properties of supersymmetry have 
allowed us to solve the problem.  


\section*{Acknowledgements} 

Fritz Rohrlich, of course, emphasized strongly that a double-well 
solution should be demonstrated.  
Also, while taking a Humboldt at the Abteiling
f\"ur Quantenphysik at the University of Ulm, two Diploma and later PhD
students, Stefan Kienle and Frank Straub, pointed out to me the difficulty
with the linear potential and kept asking me how it could be overcome.
Finally, after giving a talk on volcano potentials, Peter Milonni reminded
me of the interaction with Fritz, which finally brought this all together.   
I also thank D. Rodney Truax for many helpful comments.  
This work was supported by the US DOE.    



\end{document}